\ifdefined\XeTeXversion
\DeclareFontFamily{TU}{ptm}{}
\DeclareFontShape{TU}{ptm}{m}{n}{<-> "Times\space New\space Roman:mapping=tex-text;"}{}
\DeclareFontShape{TU}{ptm}{m}{it}{<-> "Times\space New\space Roman\space Italic:mapping=tex-text;"}{}
\DeclareFontShape{TU}{ptm}{m}{sl}{<-> "Times\space New\space Roman\space Italic:mapping=tex-text;"}{}
\DeclareFontShape{TU}{ptm}{b}{n}{<-> "Times\space New\space Roman\space Bold:mapping=tex-text;"}{}
\DeclareFontShape{TU}{ptm}{bx}{n}{<-> "Times\space New\space Roman\space Bold:mapping=tex-text;"}{}
\DeclareFontShape{TU}{ptm}{b}{it}{<-> "Times\space New\space Roman\space Bold\space Italic:mapping=tex-text;"}{}
\DeclareFontShape{TU}{ptm}{bx}{it}{<-> "Times\space New\space Roman\space Bold\space Italic:mapping=tex-text;"}{}
\DeclareFontShape{TU}{ptm}{m}{sc}{<-> "Times\space New\space Roman:mapping=tex-text;"}{}
\fi
\documentclass[journal]{IEEEtran}
\usepackage{cite}
\usepackage{graphicx}
\usepackage{multirow}
\graphicspath{{./}}
\usepackage[cmex10]{amsmath}
\usepackage{amsmath,amsfonts,amssymb}
\usepackage{subfigure}
\usepackage{amsthm}
\usepackage{algorithm}
\usepackage{stfloats}
\usepackage{bm}
\usepackage{color}
\DeclareRobustCommand{\responsecurrentchange}[1]{{\color{black}#1}}
\DeclareRobustCommand{\responsechange}[1]{{\let\currentchange\responsecurrentchange\color{black}#1}}
\DeclareRobustCommand{\currentchange}[1]{{\color{black}#1}}
\usepackage{booktabs}
\usepackage{makecell}
\usepackage{array}
\usepackage{mathrsfs}
\usepackage{url}
\usepackage{epstopdf}
\usepackage{balance}
\DeclareMathAlphabet{\mathbbold}{U}{bbold}{m}{n}
\begin{document}
\ifdefined\pdfminorversion\pdfminorversion=6\fi
\newtheorem{lemma}{Lemma}
\newtheorem{corol}{Corollary}
\newtheorem{theorem}{Theorem}
\newtheorem{proposition}{Proposition}
\newtheorem{definition}{Definition}
\newcommand{\e}{\begin{equation}}
\newcommand{\ee}{\end{equation}}
\newcommand{\eqn}{\begin{eqnarray}}
\newcommand{\eeqn}{\end{eqnarray}}
\newcolumntype{M}[1]{>{\centering\arraybackslash}m{#1}}

\title{Deep Learning-Based Tri-Hybrid Multi-User MIMO Precoding: The Blessing of  EM-Reconfigurable Antennas}

\author{Zhen Gao,~\IEEEmembership{Senior Member,~IEEE}, Kaijun Feng, Jiaxin He, Hongrui Yu, Anwen Liao, Ziwei Wan,~\IEEEmembership{Member,~IEEE}, Zhaocheng~Wang,~\IEEEmembership{Fellow,~IEEE}
        \thanks{Zhen Gao and Kaijun Feng are with the School of Information and Electronics, Beijing Institute of Technology (BIT), Beijing 100081, China (email: gaozhen16@bit.edu.cn; 3220242906@bit.edu.cn).}
        \thanks{Jiaxin He is with the School of Information and Electronics, BIT (Zhuhai), Zhuhai 519088, China (email: alvahe@bit.edu.cn).}
        \thanks{Hongrui Yu and Zhaocheng Wang are with the Department of Electronic Engineering, Tsinghua University, Beijing, 100084, China (email: yuhr23@mails.tsinghua.edu.cn; zcwang@tsinghua.edu.cn).}
        \thanks{Anwen Liao is with the School of Information and Communication, Guilin University of Electronic Technology (GUET), Guilin 541004, China  (e-mail: liaoanwen@guet.edu.cn)}
        \thanks{Ziwei Wan is with the Yangtze Delta Region Academy, BIT (Jiaxing), Jiaxing 314019, China, and also with the School of Information and Electronics, Advanced Research Institute of Multidisciplinary Sciences, BIT, Beijing 100081, China (email: ziweiwan@bit.edu.cn).}
        }


\maketitle

\begin{abstract}
Electromagnetic (EM)-reconfigurable antennas provide multiple candidate radiation patterns per element, thereby introducing an additional EM-domain degree of freedom. Integrating radiation-pattern reconfigurability, realized as EM-domain precoding, with conventional hybrid analog--digital precoding yields tri-hybrid multiple-input multiple-output (MIMO) precoding, which can substantially improve the spectral efficiency of wideband multi-user MIMO orthogonal frequency-division multiplexing (OFDM) systems. However, the joint design of EM, analog, and digital precoding remains challenging. To address this challenge, we propose a tri-hybrid precoding network (Tri-PNet) based on Conformer, an emerging neural architecture that combines the local modeling strength of convolutional neural networks with the global dependency modeling of Transformers. Furthermore, two representative radiation-pattern modes, i.e., the non-regular mode and the 3rd Generation Partnership Project (3GPP) Technical Report (TR) 38.901 mode, are investigated. Tri-PNet is trained in an unsupervised manner to jointly learn EM, analog, and digital precoding by maximizing the average sum spectral efficiency. Its radiation-pattern selection network (RPSNet) employs a Conformer encoder to capture both local and global frequency-domain correlations, whereas its hybrid analog--digital precoding network (HPNet) combines cross-attention and dual-path processing with singular-value-decomposition (SVD) and zero-forcing (ZF) priors. Simulation results under both radiation-pattern modes demonstrate that Tri-PNet outperforms random EM precoding and conventional hybrid MIMO without EM precoding, approaches the greedy EM precoding search scheme with substantially lower online complexity, and remains robust to imperfect channel state information (CSI).
\end{abstract}


\enlargethispage{\baselineskip}
\section{Introduction}

\currentchange{As 6G evolves toward brand-new application scenarios such as ubiquitous connectivity \cite{nearspace24,nearspace25,csgrant,csadaptive} and integrated sensing, communication, and computing \cite{litsp,embodied,ocdm}, high spectral efficiency under stringent hardware constraints becomes increasingly important.} Various new methodologies \cite{pengsat,pengswipt,zhangmilac} have attracted wide attention. As one of the backbone techniques, massive multiple-input multiple-output (MIMO) can substantially improve spectral efficiency through spatial multiplexing and multi-user precoding. However, a fully-digital massive MIMO transmitter requires one radio-frequency (RF) chain per antenna, causing \responsechange{circuit power consumption and hardware cost} to scale rapidly with the array size. Hybrid analog--digital precoding reduces the number of RF chains by combining a high-dimensional analog precoder with low-dimensional digital precoders \currentchange{\cite{ddl,adl,holographic}}. Nevertheless, in fully-connected hybrid MIMO architectures with hundreds or more antennas, the phase-shifter network and analog interconnection still incur substantial power consumption and implementation cost. On the other hand, restricting the array to only tens of antennas limits the achievable spatial-multiplexing and beamforming gains. This motivates the exploration of new hardware architectures that can deliver high spectral efficiency without prohibitive hardware cost.

To address this dilemma, flexible hardware architectures that introduce new physical-layer degrees of freedom have attracted increasing attention. Movable antenna (MA) systems exploit favorable propagation conditions by adjusting antenna positions \currentchange{\cite{maai,dlma,sunma}. Fluid antenna systems} (FAS) adapt to time-varying environments by selecting among candidate ports \cite{fas}, with applications including index modulation \currentchange{\cite{fae}. Rotatable antenna} (RA) systems \responsechange{reorient their boresight} \cite{rae}. These architectures enrich the spatial degrees of freedom beyond those available to conventional fixed-position arrays. \responsechange{However, mechanically actuated MA and RA implementations require continuous position or orientation adjustment, which can be difficult to accommodate in low-latency wideband multi-user transmission.}

\responsechange{Electronically switched FAS implementations can avoid continuous mechanical movement.} \responsechange{Nevertheless, such port selection imposes stringent requirements} on switching speed and reliable channel acquisition when a large set of candidate port states is available. Furthermore, both MAs and FAS exploit favorable propagation conditions by optimizing antenna positions to combat spatially selective fading. In wideband multi-carrier systems such as orthogonal frequency-division multiplexing (OFDM), the spatial fading profile varies across subcarriers, yielding distinct optimal antenna positions for different subcarriers. \responsechange{Consequently, a single static antenna position or selected FAS port must balance the channel conditions across subcarriers, which couples their wideband design.}

\responsechange{As a general hardware-agnostic concept, EM-reconfigurable antennas keep the physical position of each antenna element fixed while dynamically reshaping EM-domain radiation patterns and directly regulating the interaction between EM-domain radiated signals and the propagation environment.} Multiple practical hardware realizations exist for such EM-reconfigurable antennas, including the fluid-metal-enabled EM-reconfigurable fluid antenna system (ER-FAS) \cite{erf} and reconfigurable parasitic pixel-antenna implementations \cite{rmm}.
Taking pixel-based reconfigurable antennas as a representative example, the ON/OFF states of positive intrinsic negative (PIN)-diodes on the parasitic pixel surface are tuned to modulate the surface-current distribution, generating radiation patterns with diverse main-lobe pointing directions and half-power beamwidths (HPBWs). As a promising technology, EM-reconfigurable antennas can optimize signal transmission and reception without consuming additional spectrum resources, enlarging the physical aperture, or increasing the number of RF chains \cite{rmm}. \responsechange{However, with $N_{\rm p}$ candidates per element, the $N_{\rm t}$-element array has $(N_{\rm p})^{N_{\rm t}}$ possible pattern configurations (e.g., $N_{\rm t}=32$, $N_{\rm p}=4$, and $(N_{\rm p})^{N_{\rm t}}=4^{32}$), and each configuration changes the channel for analog and digital precoding. The resulting EM precoding, analog precoding, and digital precoding must therefore be optimized jointly under shared wideband hardware constraints.}

Recently, research on EM-reconfigurable antennas has expanded from antenna-level EM-reconfiguration hardware design to communication-oriented channel modeling and signal-processing algorithm development. For a clear comparative overview, Table~I summarizes representative MIMO architectures across the digital, analog, EM, and spatial-position reconfiguration domains, together with their implementation mechanisms. \responsechange{In the formulations compared here, tri-domain MIMO combines spatial reconfiguration with EM and digital processing, whereas the tri-hybrid setting adopted in this work combines EM, analog, and digital processing at fixed antenna positions.} Representative prior works cover both antenna-hardware prototyping and communication-algorithm exploration.

Specifically, \cite{sbo} employs successive exhaustive Boolean optimization (SEBO) to tune the pixel-connection states of a planar pixel antenna, thereby manipulating the surface-current distribution and radiation responses; \cite{acd} develops an antenna-coding mechanism for multi-user transmission, such that different coding states correspond to distinguishable radiation responses; \responsechange{\cite{ermod} proposes an FAS-based EM-reconfigurable antenna, where reshaping the liquid-metal parasitic structure reconfigures the radiation response of a fixed-position element}; \cite{rema} constructs discrete equivalent radiation locations through the switching states of a reconfigurable pixel antenna, thereby realizing an electronically movable antenna array without mechanical movement. Furthermore, \cite{rpr} combines FAS with pixel-antenna pattern reconfiguration and improves link adaptability by adjusting the beam direction and HPBW. A comprehensive survey \cite{r6g} reviews reconfigurable-antenna technologies for future wireless systems, covering hardware prototypes, system architectures, and signal-processing techniques.

\begin{table*}[t]
\centering
\scriptsize
\renewcommand{\arraystretch}{1.08}
\setlength{\tabcolsep}{1.5pt}
\caption{Comparison of reconfiguration domains, mechanisms, and limitations in representative MIMO architectures.}
\label{tab:related_work}
\begin{tabular}{M{0.78in}M{0.30in}M{0.46in}M{0.46in}M{0.46in}M{0.58in}M{1.32in}M{1.18in}}
\hline
\makecell{Architecture} & \makecell{Ref.} & \makecell{Digital\\precoding} & \makecell{Analog\\precoding} & \makecell{EM\\precoding} & \makecell{Spatial\\reconfiguration} & \makecell{Reconfiguration\\mechanism} & \makecell{Comments}\\
\hline
Hybrid MIMO & \cite{ddl} & $\checkmark$ & $\checkmark$ & & & & \makecell{\responsechange{Phase-shifter and}\\\responsechange{interconnection overhead}\\\responsechange{in fully-connected arrays}} \\
\hline
\multirow{2}{*}[-14pt]{\makecell{Movable/fluid\\antenna}} & \cite{fas,fae} & & & & \makecell{Port selection\\and index\\modulation} & & \multirow{2}{*}[-11pt]{\makecell{Low antenna-aperture\\utilization and slow\\reconfiguration response}} \\
\cline{2-7}
& \cite{maai,sunma} & $\checkmark$ & & & \makecell{Antenna\\position\\optimization} & & \\
\hline
\multirow{3}{*}[-32pt]{\makecell{EM-reconfigurable\\MIMO}} & \cite{rmm} & $\checkmark$ & & $\checkmark$ & & \makecell{EM: discrete non-regular\\radiation-pattern\\selection} & \multirow{3}{*}[-32pt]{\makecell{Requires\\low-complexity EM\\precoding design}} \\
\cline{2-7}
& \cite{rpr} & & $\checkmark$ & $\checkmark$ & & \makecell{EM: discrete 3GPP TR 38.901\\radiation-pattern selection\\with different directions\\and HPBWs} & \\
\cline{2-7}
& \cite{tcom} & $\checkmark$ & & $\checkmark$ & & \makecell{EM: joint EM-domain\\precoding and channel\\estimation} & \\
\hline
\multirow{1}{*}{Tri-domain MIMO} & \cite{tdm} & $\checkmark$ & & $\checkmark$ & & \makecell{EM and spatial:\\continuous radiation-pattern\\design and continuous\\antenna position} & \makecell{Low antenna-aperture\\utilization and slow\\reconfiguration response} \\
\hline
\multirow{3}{*}[-25pt]{Tri-hybrid MIMO} & \cite{thm} & $\checkmark$ & $\checkmark$ & $\checkmark$ & & \makecell{EM: discrete\\radiation-pattern selection\\with pixel-pattern\\switching} & \multirow{2}{*}[-10pt]{\makecell{Requires\\low-complexity\\tri-hybrid\\precoding}} \\
\cline{2-7}
& \cite{thp} & $\checkmark$ & $\checkmark$ & $\checkmark$ & & \makecell{EM: continuous\\radiation-pattern design\\with spherical-harmonic\\decomposition} & \\
\cline{2-8}
& Our work & $\checkmark$ & $\checkmark$ & $\checkmark$ & & \makecell{EM: discrete\\radiation-pattern\\selection\\with pixel-pattern\\switching} & \makecell{Low complexity and\\high spectral efficiency} \\
\hline
\end{tabular}
\\[1mm]
\parbox{0.98\textwidth}{\scriptsize\emph{Note.} In the reconfiguration mechanism column, HPBW denotes half-power beamwidth.}
\end{table*}

At the precoding level, recent studies have begun to jointly exploit the EM, analog, and digital degrees of freedom in EM-reconfigurable antenna-aided MIMO systems. The work in \cite{tcom} develops EM-domain precoding and channel-estimation techniques for reconfigurable massive MIMO, establishing a communication-theoretic basis for exploiting EM reconfigurability. The work in \cite{tdm} studies tri-domain multi-user MIMO precoding and channel estimation with a spatial-EM reconfigurable antenna. The work in \cite{thm} first investigates a tri-hybrid MIMO architecture enabled by reconfigurable antennas, revealing the potential performance gains brought by EM-domain precoding. Building upon this architecture, the work in \cite{thp} represents EM-reconfigurable radiation patterns through spherical-harmonic expansion and jointly optimizes reconfigurable radiation patterns-based EM precoders, analog precoders, and digital precoders via alternating optimization. In a different modeling approach, the work in \cite{thr} formulates EM-domain beamforming as radiation-center selection and employs a triple-loop iterative method to jointly improve spectral and energy efficiencies. Despite their demonstrated benefits, these optimization-based methods treat radiation-pattern selection (also referred to as EM precoding in this paper), analog precoding, and digital precoding as tightly coupled and require repeated updates of strongly interdependent variables. Consequently, they incur a substantial online computational burden in wideband multi-user systems, where per-subcarrier channel variation further amplifies the iteration overhead \cite{thp,thr}.

Recent advances in deep learning offer a promising alternative to iterative optimization for the coupled problems \cite{wangnoma,jiaaccess,jiasym}. In particular, \cite{prh} proposes a Transformer-based pattern-reconfigurable hybrid beamforming network (PR-HBFNet). The proposed PR-HBFNet cascades a Transformer-driven pattern-reconfigurable module for antenna-wise radiation-pattern selection with a model-driven hybrid beamforming module for hybrid analog--digital precoding. While PR-HBFNet substantially reduces the online complexity of radiation-pattern search, two limitations remain. First, its standard Transformer architecture prioritizes global dependency modeling and does not explicitly exploit the strong local correlation among adjacent subcarriers in wideband channels. \responsechange{Second, pattern selection changes the effective channel used by both hybrid-precoding branches, so the SVD analog and ZF digital priors must be refined according to this selected channel. This motivates HPNet's gated cross-attention refinement of the SVD prior and dual-path refinement of the ZF-based digital branch.} To capture both local and global frequency-domain correlations while enabling end-to-end joint optimization of pattern selection and hybrid precoding, this paper proposes a Conformer-based tri-hybrid precoding network (Tri-PNet).

By integrating convolutional local modeling with global self-attention \cite{cct}, the Conformer encoder enables the radiation-pattern selection network (RPSNet)-based EM precoding to capture both local frequency-domain structures and long-range wideband dependencies simultaneously. \responsechange{The hybrid analog--digital precoding network (HPNet) refines the SVD analog prior through gated cross-attention and the ZF-based digital output through dual-path attention, retaining the selected-channel coupling in end-to-end training.}

\currentchange{The main contributions of this work are summarized below.}
\begin{itemize}
\item \textbf{Conformer-based end-to-end tri-hybrid precoding network.} We propose Tri-PNet, a Conformer-based framework that maps the sample-wise wideband channel state information (CSI) tensor end-to-end to EM, analog, and digital precoding variables. \currentchange{Its cascaded RPSNet and HPNet are jointly trained.} RPSNet uses a Conformer encoder and Gumbel-Softmax for differentiable antenna-wise radiation-pattern selection (i.e., EM precoder design), while HPNet fuses singular-value-decomposition (SVD) and zero-forcing (ZF) priors with attention modules to generate sub-connected analog and per-subcarrier digital precoders, thereby avoiding the need for iterative optimization.

\item \textbf{\responsechange{Validation under two complementary radiation-pattern families.}} \responsechange{We evaluate Tri-PNet using two prescribed candidate pattern families.} The non-regular mode generates non-uniform responses from discrete pixel-antenna switching states \cite{rmm}, whereas the 3rd Generation Partnership Project (3GPP) Technical Report (TR) 38.901 mode constructs a structured pattern set parameterized by azimuth, elevation, and HPBW \cite{rpr}. \responsechange{These experiments establish performance for the two tested families; applicability to arbitrary antenna hardware and transfer of one trained model to a new codebook without retraining are not established.}

\item \textbf{Comprehensive performance, robustness, and complexity evaluation.} We conduct extensive simulations for wideband multi-user downlink transmission \currentchange{from the following three perspectives.} \currentchange{(i) In terms of spectral efficiency, Tri-PNet} approaches greedy-search performance under varied transmit powers, user numbers, and pattern conditions, requiring only a single forward pass. \currentchange{(ii) Regarding robustness, it maintains} advantages under HPBW variation and imperfect CSI. \currentchange{(iii) For complexity, quantified} parameter counts and floating-point operations (FLOPs) confirm a favorable trade-off of moderate offline training cost for significantly reduced online latency.

\end{itemize}

\textit{Notation:} Bold lower-case and upper-case letters denote vectors and matrices, respectively. The operators $(\cdot)^{\rm T}$, $(\cdot)^{\rm H}$, and $(\cdot)^*$ denote transpose, conjugate transpose, and conjugate, respectively. The Frobenius norm is denoted by $\|\cdot\|_{\rm F}$. The circularly symmetric complex Gaussian distribution with variance $\sigma^2$ is denoted by $\mathcal{CN}(0,\sigma^2)$. The operators $\Re\{\cdot\}$ and $\Im\{\cdot\}$ denote \currentchange{the real and imaginary parts, respectively.} $\mathbb{E}\{\cdot\}$ denotes expectation; $\mathrm{vec}(\cdot)$ and $\mathrm{reshape}(\cdot)$ denote \currentchange{vectorization and the corresponding inverse operation}, \currentchange{respectively. The operator $\odot$ denotes} the Hadamard product.

\section{System Model and Problem Formulation }

We consider a downlink multi-user MIMO system where a base station (BS) serves $U$ single-antenna users over an OFDM waveform with $N_{\rm c}$ subcarriers. The BS is equipped with a uniform planar array containing $N_{\rm t}=N_xN_z$ radiation-pattern-reconfigurable (RPR) antennas and $N_{\rm RF}$ RF chains, with $U\leq N_{\rm RF}\ll N_{\rm t}$. Here, $N_x$ and $N_z$ denote the numbers of antennas along the horizontal and vertical directions, respectively. To exploit the EM-domain degree of freedom, each antenna can select one radiation pattern from a set of $N_{\rm p}$ candidate EM radiation patterns.

\begin{figure}[t]
\centering
\includegraphics[width=0.96\linewidth,height=0.74\linewidth]{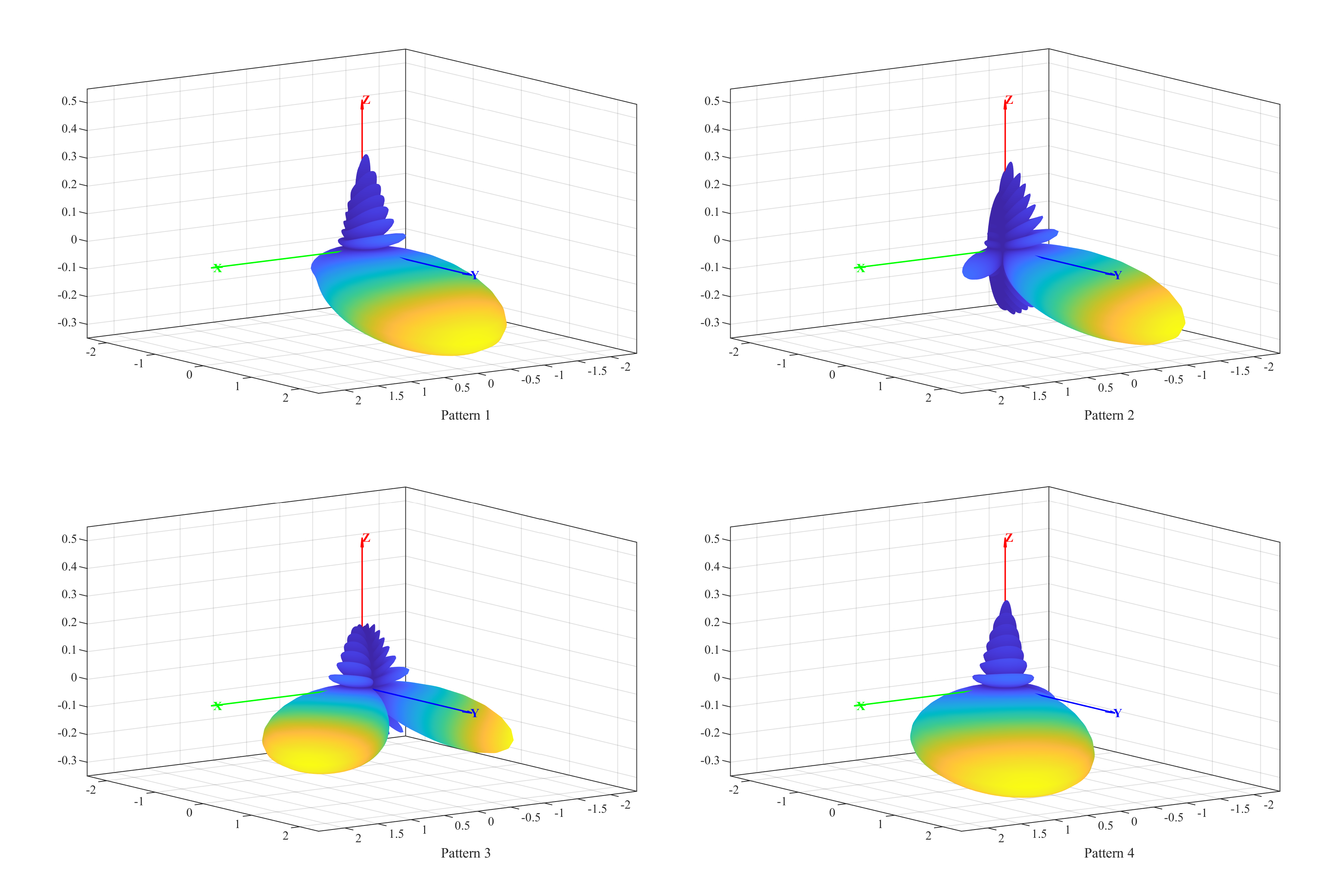}
\caption{Four Huawei-provided, physically implementable basis radiation patterns used for the non-regular mode \cite{rmm}. For each antenna element, the horizontal, vertical, and normal directions are the $x$, $z$, and $y$ axes, respectively.}
\label{fig:rmm_patterns}
\end{figure}

\begin{figure*}[t]
\centering
\includegraphics[width=0.90\textwidth,keepaspectratio]{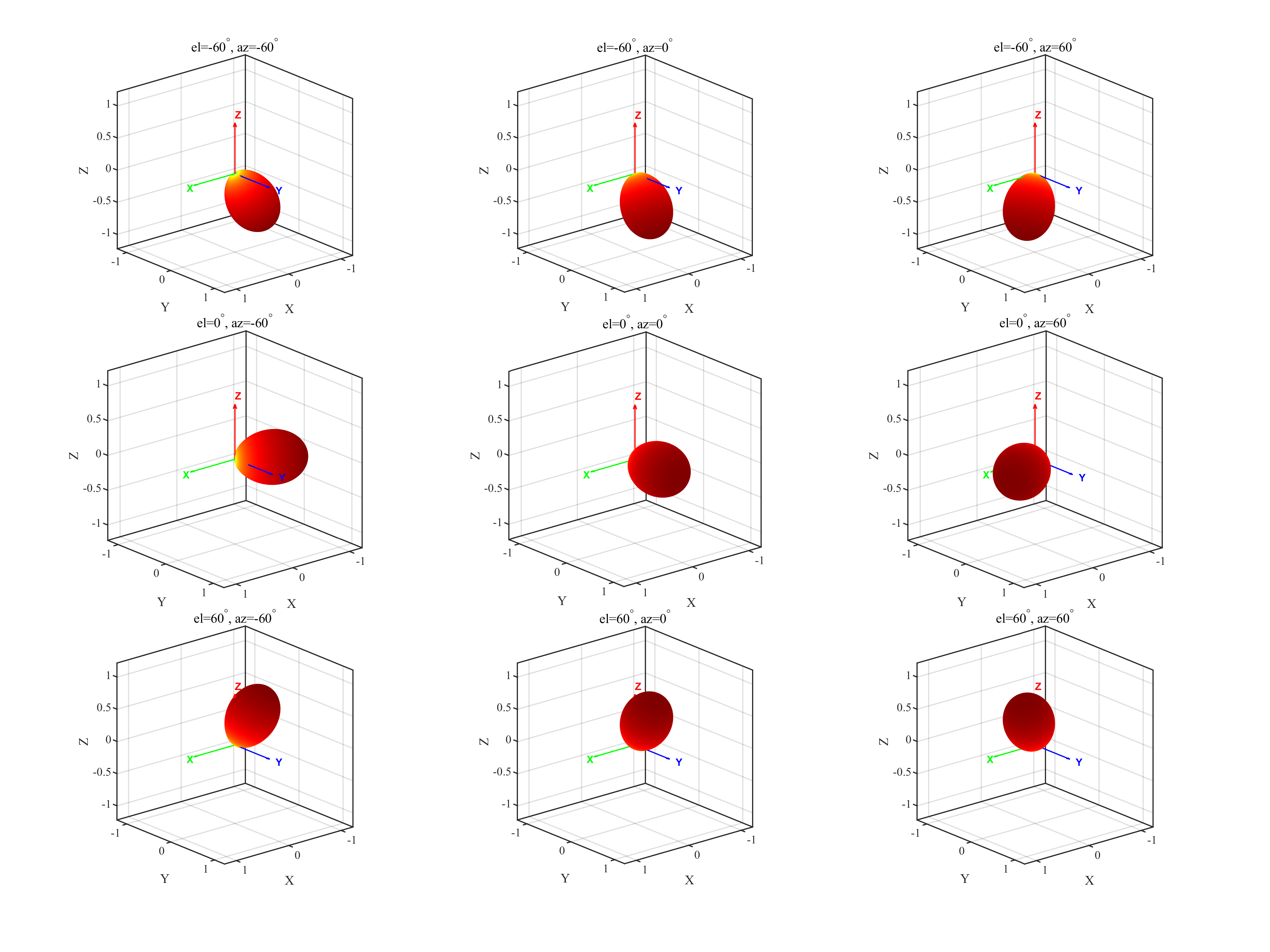}
\caption{Nine candidate radiation patterns of the 3GPP TR 38.901 mode adopted in \cite{rpr}. With HPBW $=90^\circ$, the elevation and azimuth main-lobe pointing directions are sampled at $-60^\circ$, $0^\circ$, and $60^\circ$. For each antenna element, the horizontal, vertical, and normal directions are the $x$, $z$, and $y$ axes.}
\label{fig:rpr_patterns}
\end{figure*}

 \currentchange{We consider two EM radiation-pattern modes, namely the non-regular mode} \cite{rmm}\footnote{The four non-regular radiation patterns in Fig.~\ref{fig:rmm_patterns} are the Huawei-provided basis patterns reported as physically implementable in \cite{rmm,prce}.} and the 3GPP TR 38.901 mode \cite{rpr}. The non-regular mode follows the EM-reconfigurable pixel-antenna design in \cite{rmm}, where the ON/OFF configurations of PIN-diode switches connecting the conductive pixels are reconfigured so that different radiation directions and pattern shapes can be realized. The 3GPP TR 38.901 mode follows the directional radiation-pattern construction in \cite{rpr}, which adjusts the elevation and azimuth angles of the main-lobe direction and the HPBW of the reference element pattern. An example of the 3GPP TR 38.901 mode is shown in Fig.~\ref{fig:rpr_patterns}, where HPBW $=90^\circ$ is adopted and both axes are sampled at $-60^\circ$, $0^\circ$, and $60^\circ$, yielding $3\times3=9$ overlapping radiation patterns with distinct main-lobe orientations. Although the two modes stem from different physical mechanisms, both are characterized by the scalar complex field gain $G_p^{\rm EM}(\phi,\theta)$ imposed upon the corresponding propagation path.
\begin{figure}[t]
\centering
\includegraphics[width=0.95\linewidth]{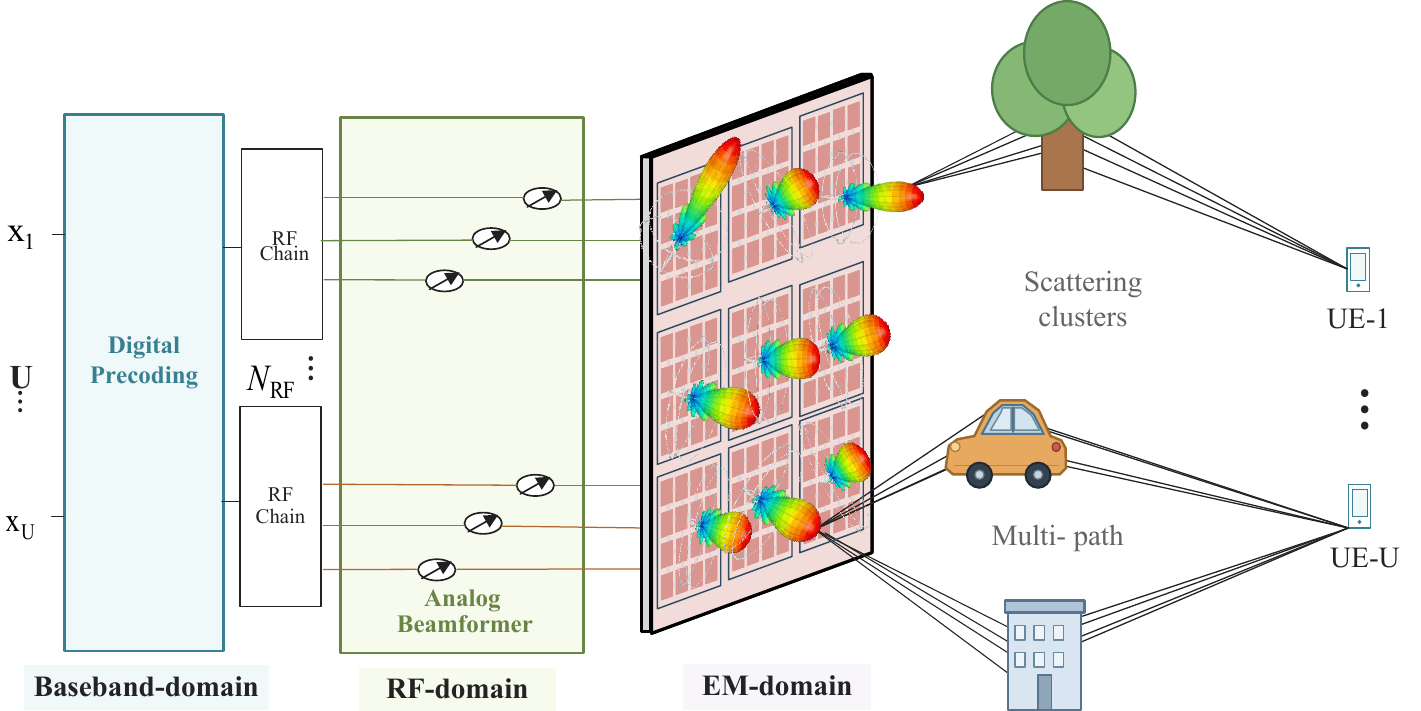}
\caption{\responsechange{System model of the proposed tri-hybrid multi-user downlink transmitter. The baseband processor applies digital precoding, the sub-connected RF phase-shift network applies analog precoding, and the EM-reconfigurable antennas radiate using their selected patterns.}}
\label{fig:reconfigurable_antenna}
\end{figure}

For the $u$-th user, let $L_u$ denote the number of propagation paths. The corresponding reconfigurable EM-domain gain matrix is defined as
\begin{equation}
\mathbf{G}_{u}^{\rm EM}
\in\mathbb{C}^{N_{\rm p}\times L_u},\quad
[\mathbf{G}_{u}^{\rm EM}]_{p,\ell}
=G_p^{\rm EM}(\phi_{u,\ell},\theta_{u,\ell}),
\label{eq:EM_matrix}
\end{equation}
where $p\in\{1,\ldots,N_{\rm p}\}$ indexes the row and the corresponding candidate radiation pattern, and $\ell\in\{1,\ldots,L_u\}$ indexes the column and the corresponding propagation path. Thus, the $(p,\ell)$-th entry represents the complex field gain of the $p$-th radiation pattern along the $\ell$-th propagation path of the $u$-th user. When a source pattern is specified in the power-gain domain, normalization is performed prior to conversion to the corresponding field-domain response. This enables direct utilization of \(G_{p}^{\rm EM}(\phi, \theta)\) for channel computation.

For the planar array, the antenna index $n$ is mapped to the planar coordinate $(n_x,n_z)$, where $n_x=0,\ldots,N_x-1$ and $n_z=0,\ldots,N_z-1$. According to \cite{rpr}, the transmit array response of the $(n_x,n_z)$-th antenna for the path with elevation angle $\phi$ and azimuth angle $\theta$ is written as
\begin{equation}
a_{n_x,n_z}(\phi,\theta)
=\frac{1}{\sqrt{N_{\rm t}}}
e^{j\frac{2\pi d}{\lambda}
\left(n_x\cos\phi\cos\theta+n_z\cos\phi\sin\theta\right)},
\label{eq:upa_response}
\end{equation}
where $d=\lambda/2$ is the antenna spacing and $\lambda$ is the carrier wavelength. Here, $\phi$ is the elevation angle measured from the array plane, whereas $\theta$ is the azimuth angle; the $\cos\phi$ terms in \eqref{eq:upa_response} follow this elevation-angle convention.

With this notation, let $f_q$ denote the subcarrier frequency of the $q$-th subcarrier. The associated channel coefficient for the $u$-th user, the $p$-th pattern, and the $n$-th antenna can be written as
\begin{equation}
\begin{aligned}
h_{p,q,u,n}
&=
\sum_{\ell=1}^{L_u}
\alpha_{u,\ell}
G_p^{\rm EM}(\phi_{u,\ell},\theta_{u,\ell})\\
&\quad\times
a_{n_x,n_z}^*(\phi_{u,\ell},\theta_{u,\ell})
e^{-j2\pi f_q\tau_{u,\ell}},
\end{aligned}
\label{eq:unified_channel}
\end{equation}
where $\alpha_{u,\ell}$, $\tau_{u,\ell}$, $\phi_{u,\ell}$, and $\theta_{u,\ell}$ denote the complex path gain, delay, elevation angle, and azimuth angle of the $\ell$-th path of the $u$-th user, respectively. The conjugation of the transmit array response is consistent with the downlink channel representation in \eqref{eq:unified_channel}. Therefore, different radiation patterns associated with the same $n$-th antenna share the same wideband channel expression and differ only in the construction of $G_p^{\rm EM}(\phi,\theta)$.

Here, we define the radiation pattern selection matrix, also referred to as the EM precoding matrix, as
\begin{equation}
\mathbf{F}_{\rm EM}\in\{0,1\}^{N_{\rm t}\times N_{\rm p}}, \quad
\sum_{p=1}^{N_{\rm p}}[\mathbf{F}_{\rm EM}]_{n,p}=1,\ \forall n.
\end{equation}
\responsechange{If $[\mathbf{F}_{\rm EM}]_{n,p}=1$, the $n$-th antenna selects the $p$-th radiation pattern as the EM precoder. }\responsechange{\currentchange{This constraint applies row-wise, and different antennas may select the same pattern index.}} Equivalently, the selected radiation pattern index of the $n$-th antenna is denoted by $\rho_n\in\{1,2,\ldots,N_{\rm p}\}$ and $n\in\{1,2,\ldots,N_{\rm t}\}$.

Given $\mathbf{F}_{\rm EM}$, let $\mathbf{h}_{u}^{\rm EM}[q]\in\mathbb{C}^{N_{\rm t}}$ denote the channel vector after EM precoding from all $N_t $ antennas to the $u$-th user. The channel gain associated with the $n$-th antenna can also be expressed as
\begin{equation}
[\mathbf{h}_{u}^{\rm EM}[q]]_{n}
=\sum_{p=1}^{N_{\rm p}}[\mathbf{F}_{\rm EM}]_{n,p}
h_{p,q,u,n}
=h_{\rho_n,q,u,n}.
\label{eq:selected_channel}
\end{equation}
And
\begin{equation}
\begin{aligned}
\mathbf{h}_{u}^{\rm EM}[q]
&=
\left[
h_{\rho_1,q,u,1},
h_{\rho_2,q,u,2},
\ldots,
h_{\rho_{N_{\rm t}},q,u,N_{\rm t}}
\right]^{\rm T}.
\end{aligned}
\end{equation}
Thus, selecting a pattern changes the complex field gain of the corresponding propagation paths for each transmit antenna and thereby forms the pattern-aware effective channel. Therefore, the channel matrix given the EM precoding matrix $\mathbf{F}_{\rm EM}$ on the $q$-th subcarrier is then defined as
\begin{equation}
\mathbf{H}^{\rm EM}[q]
=\left[
\mathbf{h}_{1}^{\rm EM}[q],
\ldots,
\mathbf{h}_{U}^{\rm EM}[q]\right]^{\rm T}
\in\mathbb{C}^{U\times N_{\rm t}}.
\label{eq:hpat_matrix}
\end{equation}
The downlink received signal of the $u$-th user on the $q$-th subcarrier is
\begin{equation}
y_u[q]=
\left(\mathbf{h}_{u}^{\rm EM}[q]\right)^{\rm T}
\mathbf{F}_{\rm RF}\mathbf{F}_{\rm BB}[q]\mathbf{s}[q]
+z_u[q],
\label{eq:received_signal}
\end{equation}
where $\mathbf{F}_{\rm RF}\in\mathbb{C}^{N_{\rm t}\times N_{\rm RF}}$ denotes the sub-connected analog precoding matrix, $\mathbf{F}_{\rm BB}[q]\in\mathbb{C}^{N_{\rm RF}\times U}$ denotes the digital precoding matrix on the $q$-th subcarrier, $\mathbf{s}[q]=[s_1[q],\ldots,s_U[q]]^{\rm T}\in\mathbb{C}^{U}$ is the data symbol vector satisfying $\mathbb{E}\{\mathbf{s}[q]\mathbf{s}^{\rm H}[q]\}=\mathbf{I}_{U}$, $s_u[q]$ denotes the data symbol intended for the $u$-th user, and $z_u[q]\sim\mathcal{CN}(0,\sigma^2)$ is the additive white Gaussian noise. The EM precoding matrix $\mathbf{F}_{\rm EM}$ determines $\mathbf{H}^{\rm EM}[q]$, based on which $\mathbf{F}_{\rm RF}$ determines the RF phase-shift network and $\mathbf{F}_{\rm BB}[q]$ performs subcarrier-wise digital-domain multi-user precoding. The transmit power constraint is
\begin{equation}
\left\|\mathbf{F}_{\rm RF}\mathbf{F}_{\rm BB}[q]\right\|_{\rm F}^{2}
\leq P_{\rm t}, \quad q=1,\ldots,N_{\rm c}.
\end{equation}

Let $\mathbf{f}_{v}[q]$ be the $v$-th column of
$\mathbf{F}[q]=\mathbf{F}_{\rm RF}\mathbf{F}_{\rm BB}[q]$. The signal-to-interference-plus-noise ratio (SINR) of the $u$-th user on the $q$-th subcarrier is
\begin{equation}
\gamma_u[q]=
\frac{\left|\left(\mathbf{h}_{u}^{\rm EM}[q]\right)^{\rm T}\mathbf{f}_u[q]\right|^2}
{\sigma^2+\sum_{v=1,v\neq u}^{U}
\left|\left(\mathbf{h}_{u}^{\rm EM}[q]\right)^{\rm T}\mathbf{f}_v[q]\right|^2},
\end{equation}
where $\sigma^2$ denotes the variance of the additive white Gaussian noise. The average sum spectral efficiency is
\begin{equation}
R=
\frac{1}{N_{\rm c}}\sum_{q=1}^{N_{\rm c}}\sum_{u=1}^{U}
\log_2\left(1+\gamma_u[q]\right).
\label{eq:se}
\end{equation}
Let $\mathcal{S}_r$ denote the antenna index set connected to the $r$-th RF chain. The goal is to maximize $R$ by jointly designing $\mathbf{F}_{\rm EM}$, $\mathbf{F}_{\rm RF}$, and $\{\mathbf{F}_{\rm BB}[q]\}_{q=1}^{N_{\rm c}}$, which can be formulated as
\begin{equation}
\begin{aligned}
\max_{\mathbf{F}_{\rm EM},\mathbf{F}_{\rm RF},\{\mathbf{F}_{\rm BB}[q]\}_{q=1}^{N_{\rm c}}}
&\quad R\\
\mathrm{s.t.}
&\quad \mathbf{F}_{\rm EM}\in\{0,1\}^{N_{\rm t}\times N_{\rm p}},\\
&\quad \sum_{p=1}^{N_{\rm p}}[\mathbf{F}_{\rm EM}]_{n,p}=1,\quad \forall n,\\
&\quad [\mathbf{F}_{\rm RF}]_{n,r}=0,\quad n\notin\mathcal{S}_{r},\\
&\quad |[\mathbf{F}_{\rm RF}]_{n,r}|=1,\quad n\in\mathcal{S}_{r},\\
&\quad \left\|\mathbf{F}_{\rm RF}\mathbf{F}_{\rm BB}[q]\right\|_{\rm F}^{2}\leq P_{\rm t},
\quad \forall q.
\end{aligned}
\label{eq:optimization_problem}
\end{equation}
Equivalently, the feasible analog-precoder set consists of the sub-connected matrices whose active entries have unit modulus, whereas the feasible EM-precoder set consists of the binary one-hot pattern-selection matrices defined above. Problem~\eqref{eq:optimization_problem} is mixed-integer due to the binary-valued matrix $\mathbf{F}_{\rm EM}$ and non-convexity stemming from the constant-modulus constraint on $\mathbf{F}_{\rm RF}$. \responsechange{Both $\mathbf{F}_{\rm EM}$ and $\mathbf{F}_{\rm RF}$ are common to all considered OFDM subcarriers, so each antenna retains one selected pattern across them; only $\mathbf{F}_{\rm BB}[q]$ is subcarrier dependent. These shared EM and analog configurations couple the wideband design.} These properties motivate a direct learning-based solution that jointly infers the EM-domain, analog-domain, and digital-domain variables.

\section{Proposed Conformer-based Tri-Hybrid Multi-User MIMO Precoding}

\begin{figure*}[t]
\centering
\includegraphics[width=1\textwidth]{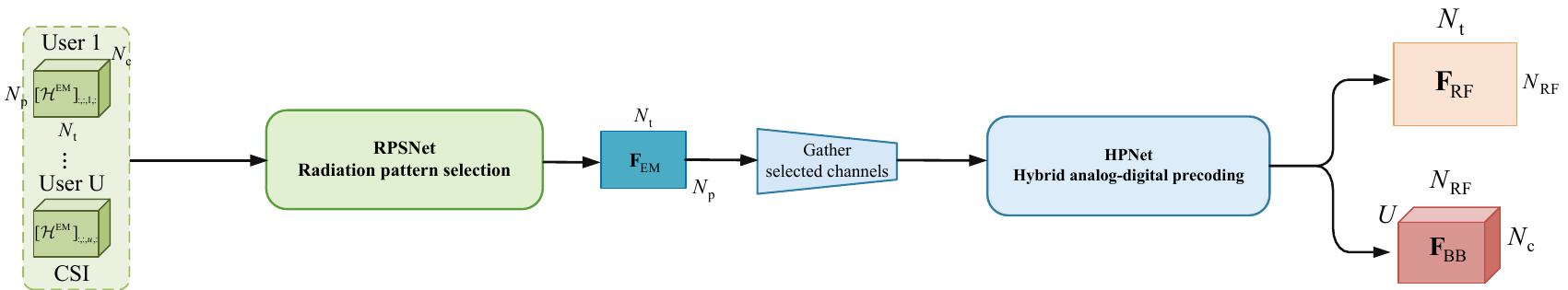}
\caption{Overview of the proposed Conformer-based Tri-PNet. RPSNet selects the antenna-wise EM precoding matrix $\mathbf{F}_{\rm EM}$, which gathers the selected channels $\mathbf{H}^{\rm EM}[q]$ for HPNet to generate $\mathbf{F}_{\rm RF}$ and $\mathbf{F}_{\rm BB}[q]$.}
\label{fig:architecture}
\end{figure*}

Following the cascaded design philosophy of learning-based hybrid analog-digital precoding, Tri-PNet maps the sample-wise wideband CSI tensor to the EM, analog, and digital precoding variables as
\begin{equation}
\left(\widehat{\mathbf{F}}_{\rm EM},\widehat{\mathbf{F}}_{\rm RF},\{\widehat{\mathbf{F}}_{\rm BB}[q]\}_{q=1}^{N_{\rm c}}\right)
=\mathcal{F}_{\Theta}(\mathcal{H}^{\rm EM}),
\end{equation}
where $\Theta$ denotes all trainable parameters of Tri-PNet and $\mathcal{F}_{\Theta}(\cdot)$ is the cascade of the RPSNet and the HPNet. The overview in Fig.~\ref{fig:architecture} shows these two cascaded modules. The input channel tensor is $\mathcal{H}^{\rm EM}\in\mathbb{C}^{N_{\rm p}\times N_{\rm c}\times U\times N_{\rm t}}$, whose $(p,q,u,n)$-th entry is
\begin{equation}
[\mathcal{H}^{\rm EM}]_{p,q,u,n}
=h_{p,q,u,n}.
\end{equation}
\responsechange{RPSNet selects one pattern per antenna from the candidate-pattern CSI tensor $\mathcal{H}^{\rm EM}$ and gathers $[\mathbf{H}^{\rm EM}[q]]_{u,n}=h_{\rho_n,q,u,n}$, yielding the selected antenna-domain channel $\mathbf{H}^{\rm EM}[q]\in\mathbb{C}^{U\times N_{\rm t}}$ in \eqref{eq:hpat_matrix}. HPNet designs $\mathbf{F}_{\rm RF}$ from these selected channels, then uses the RF-chain-domain channel $\mathbf{H}_{\rm RF}[q]=\mathbf{H}^{\rm EM}[q]\mathbf{F}_{\rm RF}\in\mathbb{C}^{U\times N_{\rm RF}}$ in \eqref{eq:hrf} to generate $\mathbf{F}_{\rm BB}[q]$ and enforce power normalization.}

\responsechange{This is the CSI-based precoder-design order shown in Fig.~\ref{fig:architecture}. The transmitted-data flow in Fig.~\ref{fig:reconfigurable_antenna} instead applies digital precoding, then analog precoding, and finally radiation through the selected-pattern antennas. The following subsections detail the two jointly trained modules.}

\subsection{RPSNet-based EM Precoding}
\begin{figure*}[t]
\centering
\includegraphics[width=0.96\textwidth]{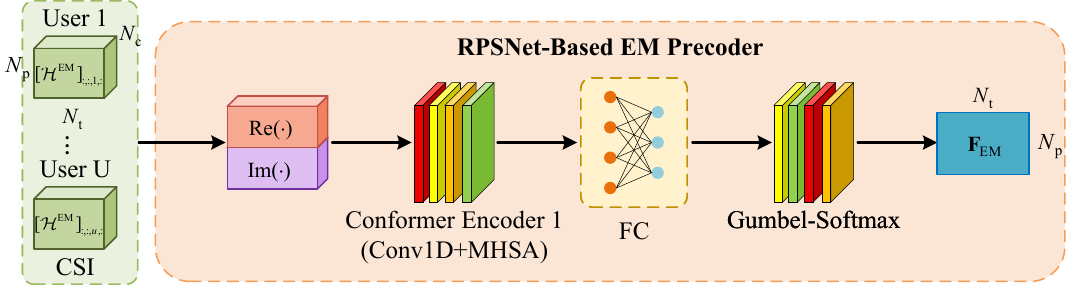}
\caption{Detailed architecture of RPSNet. The real and imaginary parts of the pattern-aware CSI are embedded and processed by a Conformer encoder to produce antenna-wise pattern logits. The Gumbel-Softmax selector generates $\mathbf{F}_{\rm EM}$, which gathers $\mathbf{H}^{\rm EM}[q]$ for $q=1,\ldots,N_{\rm c}$.}
\label{fig:rpsnet_architecture}
\end{figure*}

The RPSNet, detailed in Fig.~\ref{fig:rpsnet_architecture}, is responsible for learning the discrete radiation pattern selection matrix $\mathbf{F}_{\rm EM}$. The input CSI tensor is first rearranged along the subcarrier dimension, and the real and imaginary parts are concatenated as
\begin{equation}
\mathbf{z}_{q}
=
\left[
\Re\{\mathrm{vec}([\mathcal{H}^{\rm EM}]_{:,q,:,:})\}^{\rm T},
\Im\{\mathrm{vec}([\mathcal{H}^{\rm EM}]_{:,q,:,:})\}^{\rm T}
\right]^{\rm T},
\end{equation}
where $\mathbf{z}_{q}\in\mathbb{R}^{2N_{\rm p}UN_{\rm t}}$ denotes the feature vector of the $q$-th subcarrier. 
The sequence $\{\mathbf{z}_{q}\}_{q=1}^{N_{\rm c}}$ is projected to a latent dimension by a linear embedding layer and then processed by a Conformer encoder \cite{cct}. \responsechange{In \eqref{eq:unified_channel}, neighboring subcarriers share the path gains, delays, and angles, while their path phases differ by $e^{-j2\pi\Delta f\tau_{u,\ell}}$. When these inter-subcarrier phase changes are small, the channel exhibits local frequency correlation. The convolution acts along the subcarrier-indexed sequence $\{\mathbf{z}_{q}\}_{q=1}^{N_{\rm c}}$ in (15), rather than along antenna positions; global self-attention complements it by relating features over the full subcarrier span.} The Conformer operation can be written compactly as
\begin{equation}
\mathbf{U}_{\rm RPSNet}
=\mathcal{C}_{\rm RPSNet}
\left(
\left[\mathbf{W}_{\rm e}\mathbf{z}_{1},\ldots,
\mathbf{W}_{\rm e}\mathbf{z}_{N_{\rm c}}\right]^{\rm T}
+\mathbf{P}_{\rm e}
\right),
\end{equation}
where $\mathbf{W}_{\rm e}\in\mathbb{R}^{d_{\rm model}\times 2N_{\rm p}UN_{\rm t}}$ is the embedding matrix, $\mathbf{P}_{\rm e}\in\mathbb{R}^{N_{\rm c}\times d_{\rm model}}$ is the positional encoding, $d_{\rm model}$ is the latent dimension, $\mathbf{U}_{\rm RPSNet}\in\mathbb{R}^{N_{\rm c}\times d_{\rm model}}$ is the encoded feature, and $\mathcal{C}_{\rm RPSNet}(\cdot)$ denotes the Conformer encoder. The encoded feature is flattened and mapped to the antenna-wise pattern logits
\begin{equation}
\mathbf{A}_{\rm EM}
=\mathrm{reshape}
\left(\mathbf{W}_{\rm p}\mathrm{vec}(\mathbf{U}_{\rm RPSNet})+\mathbf{b}_{\rm p}\right)
\in\mathbb{R}^{N_{\rm t}\times N_{\rm p}}.
\end{equation}
Here, $\mathbf{W}_{\rm p}\in\mathbb{R}^{N_{\rm t}N_{\rm p}\times N_{\rm c}d_{\rm model}}$ and $\mathbf{b}_{\rm p}\in\mathbb{R}^{N_{\rm t}N_{\rm p}}$ are the output-projection parameters.

Since the radiation pattern set decision is discrete, a differentiable Gumbel-Softmax selector is used during training. The relaxed selection probability of the $n$-th antenna choosing the $p$-th pattern is
\begin{equation}
\eta_{n,p}=\frac{[\mathbf{A}_{\rm EM}]_{n,p}+g_{n,p}}{\tau_{\rm g}},
\end{equation}
\begin{equation}
[\widetilde{\mathbf{F}}_{\rm EM}]_{n,p}
=
\frac{\exp(\eta_{n,p})}{\sum_{i=1}^{N_{\rm p}}\exp(\eta_{n,i})},
\end{equation}
where $g_{n,p}$ is the Gumbel noise and $\tau_{\rm g}$ is the temperature parameter. During training at epoch $e$, the temperature follows $\tau_{\rm g}=\max\{0.1,2\exp(-0.03e)\}$, which progressively sharpens the soft selection probabilities. A straight-through hard decision is used in the forward pass so that $\mathbf{F}_{\rm EM}$ remains one-hot for each antenna, while the gradient is back-propagated through $\widetilde{\mathbf{F}}_{\rm EM}$. During inference, the selection is obtained by
\begin{equation}
\rho_n=\arg\max_{p}[\mathbf{A}_{\rm EM}]_{n,p},\quad
[\mathbf{F}_{\rm EM}]_{n,\rho_n}=1.
\end{equation}
After the radiation pattern of each antenna is selected, the corresponding entries are gathered from the original pattern-aware channel tensor according to \eqref{eq:selected_channel}, and the selected channel matrices $\{\mathbf{H}^{\rm EM}[q]\}_{q=1}^{N_{\rm c}}$ are passed to the HPNet. This gathering operation keeps the discrete radiation-pattern selection coupled with every antenna, user, and subcarrier, instead of reducing the pattern selection to a sample-level label.

\subsection{HPNet-based Hybrid Analog-Digital Precoding}
\begin{figure*}[t]
\centering
\includegraphics[width=0.96\textwidth]{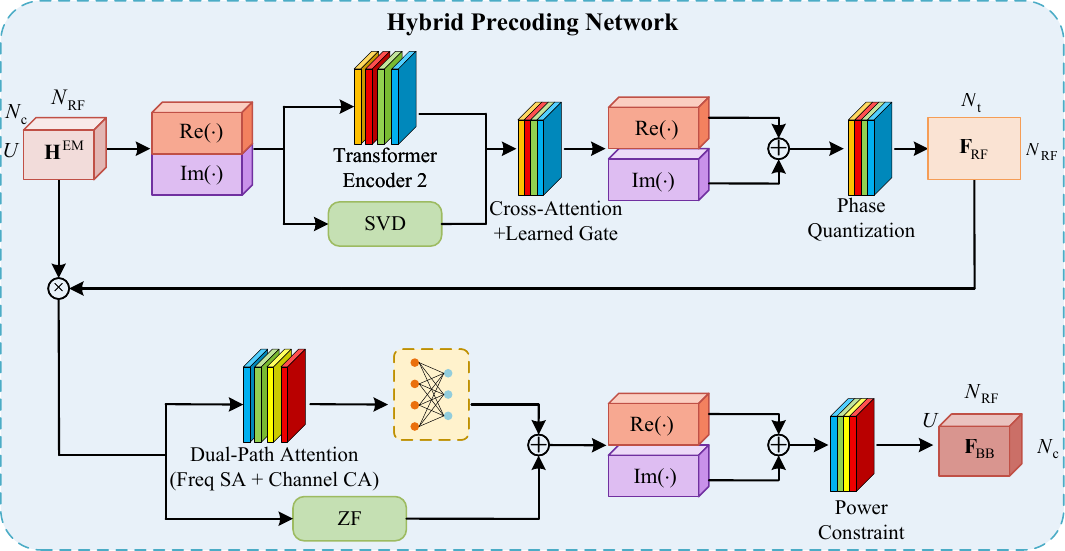}
\caption{Detailed architecture of HPNet. Its analog branch combines a Transformer channel feature with an SVD prior through gated cross-attention to obtain the quantized $\mathbf{F}_{\rm RF}$. Its digital branch processes $\mathbf{H}_{\rm RF}[q]$ using dual-path attention to produce the normalized $\mathbf{F}_{\rm BB}[q]$.}
\label{fig:hpnet_architecture}
\end{figure*}

The HPNet, detailed in Fig.~\ref{fig:hpnet_architecture}, jointly designs analog and digital precoding after radiation pattern selection. For the analog part, the network first uses $\{\mathbf{H}^{\rm EM}[q]\}_{q=1}^{N_{\rm c}}$ to construct an SVD-based analog precoder as a model-driven prior. Specifically, for each user and subcarrier, we construct the channel Gram matrix $\mathbf{C}_u[q] = (\mathbf{h}_u^{\mathrm{EM}}[q])^{*} (\mathbf{h}_u^{\mathrm{EM}}[q])^{\mathrm{T}}$, apply SVD to this matrix, and extract its dominant right singular vector.
These directions are orthogonalized, aggregated across the subcarriers, and restricted to each subarray, after which the principal subarray direction provides the corresponding RF-chain prior. Under the sub-connected RF structure, each RF chain is connected to $\frac{N_{\rm t}}{N_{\rm RF}}$ antennas. Let $\mathcal{S}_r$ denote the set of antenna indices connected to the $r$-th RF chain through phase shifters.  The analog precoder satisfies
\begin{equation}
[\mathbf{F}_{\rm RF}]_{n,r}=0,\quad n\notin\mathcal{S}_r,
\end{equation}
and only the nonzero entries with $n \in \mathcal{S}_r$ are optimized.

In the neural analog branch, the real and imaginary parts of $\mathbf{H}^{\rm EM}[q]$ are concatenated and embedded along the subcarrier dimension. A Transformer encoder extracts the pattern-aware channel feature, while the SVD-based analog precoder is projected as the query of a cross-attention module. The analog correction can be expressed as
\begin{equation}
\Delta\mathbf{F}_{\rm RF}
=
\mathcal{A}_{\rm RF}
\left(
\mathcal{T}_{\rm RF}(\mathbf{H}^{\rm EM}),
\mathbf{F}_{\rm RF}^{\rm svd}
\right),
\end{equation}
where $\mathcal{T}_{\rm RF}(\cdot)$ denotes the analog-branch Transformer encoder and $\mathcal{A}_{\rm RF}(\cdot)$ denotes the gated cross-attention refinement. Both $\mathbf{F}_{\rm RF}^{\rm svd}$ and $\Delta\mathbf{F}_{\rm RF}$ lie in $\mathbb{C}^{N_{\rm t}\times N_{\rm RF}}$. The refined analog precoder before quantization is
\begin{equation}
\widetilde{\mathbf{F}}_{\rm RF}
=\mathbf{F}_{\rm RF}^{\rm svd}
+\alpha_{\rm tr}\left(\Delta\mathbf{F}_{\rm RF}\odot\mathbf{M}_{\rm RF}\right),
\end{equation}
where $\mathbf{M}_{\rm RF}\in\{0,1\}^{N_{\rm t}\times N_{\rm RF}}$ is the sub-connected mask and $\alpha_{\rm tr}$ is a training-stage coefficient used to gradually introduce the neural correction. \currentchange{The phase shifter constraint is then enforced by finite-resolution phase quantization as}
\begin{equation}
[\mathbf{F}_{\rm RF}]_{n,r}
=
\begin{cases}
e^{j\mathcal{Q}_{b}(\angle[\widetilde{\mathbf{F}}_{\rm RF}]_{n,r})}, & n\in\mathcal{S}_r,\\
0, & n\notin\mathcal{S}_r,
\end{cases}
\label{eq:frf_quant}
\end{equation}
where $\mathcal{Q}_{b}(\cdot)$ denotes the $b$-bit phase quantizer; all simulations use $b=4$. In the implementation, a straight-through estimator is used so that the quantized phase is applied in the forward pass and the pre-quantized phase receives gradients during back-propagation.

For the digital part, the selected antenna-domain channel is grouped according to the sub-connected mapping specified by $\{\mathcal{S}_r\}^{N_{\rm RF}}_{r=1}$. After analog precoding, the RF-chain-domain effective channel on the $q$-th subcarrier is
\begin{equation}
\mathbf{H}_{\rm RF}[q]
=\mathbf{H}^{\rm EM}[q]\mathbf{F}_{\rm RF}
\in\mathbb{C}^{U\times N_{\rm RF}}.
\label{eq:hrf}
\end{equation}
In an implementation that stores antennas in subarray order, the antennas are temporarily permuted before this multiplication and the analog precoder is mapped back to the physical antenna order afterward; the mathematical operation remains \eqref{eq:hrf}. To match the digital prior, let $\mathbf{v}_{u}[q]=\left([\mathbf{H}_{\rm RF}[q]]_{u,:}\right)^{\rm H}/\left\|[\mathbf{H}_{\rm RF}[q]]_{u,:}\right\|_2$ denote the normalized conjugate-transpose of the $u$-th row of $\mathbf{H}_{\rm RF}[q]$, and define $\mathbf{V}[q]=[\mathbf{v}_{1}[q],\ldots,\mathbf{v}_{U}[q]]\in\mathbb{C}^{N_{\rm RF}\times U}$. \currentchange{A normalized-channel-based ZF digital precoder is then used as another model-driven prior, given by}
\begin{equation}
\mathbf{F}_{\rm BB}^{\rm zf}[q] = \mathbf{V}[q] \left(\mathbf{V}^{\rm H}[q]\mathbf{V}[q]\right)^{-1}.
\label{eq:zf_normalized}
\end{equation}
Note that the ZF prior in \eqref{eq:zf_normalized} differs from the standard ZF by normalizing each user channel vector to unit norm before forming $\mathbf{V}[q]$, thereby reducing the influence of unequal channel gains across users under the sub-connected architecture. Meanwhile, the real-valued representation of $\mathbf{H}_{\rm RF}[q]$ is processed by a dual-path attention block. The self-attention path captures cross-subcarrier correlation, and the cross-attention path refines the digital feature with the RF-chain-domain channel representation. The digital output is written as
\begin{equation}
\widetilde{\mathbf{F}}_{\rm BB}[q] = \mathbf{F}_{\rm BB}^{\rm zf}[q] +\alpha_{\rm tr}\mathcal{D}_{\rm BB} \left(\{\mathbf{H}_{\rm RF}[i]\}_{i=1}^{N_{\rm c}}\right)_{q},
\end{equation}
where $\mathcal{D}_{\rm BB}(\cdot)$ denotes the dual-path digital precoding branch, the subscript $q$ denotes the output corresponding to the $q$-th subcarrier, and $\alpha_{\rm tr}$ is the same training-stage coefficient used in the analog branch. \currentchange{Finally, the digital precoding matrices are normalized together with the analog precoder to satisfy the transmit power constraint as}
\begin{equation}
\mathbf{F}_{\rm BB}[q] = \frac{\sqrt{P_{\rm t}}\widetilde{\mathbf{F}}_{\rm BB}[q]} {\left\|\mathbf{F}_{\rm RF}\widetilde{\mathbf{F}}_{\rm BB}[q]\right\|_{\rm F}}, \quad q=1,\ldots,N_{\rm c}.
\label{eq:power_norm}
\end{equation}
Equation~\eqref{eq:power_norm} enforces the power constraint independently on every subcarrier, i.e., $\|\mathbf{F}_{\rm RF}\mathbf{F}_{\rm BB}[q]\|_{\rm F}^{2}=P_{\rm t}$ for all $q$. In this way, the HPNet combines model-driven beamforming priors with learnable attention modules, which improves training stability while retaining the flexibility of deep learning-based precoding.

\subsection{Training Objective}
The whole network is trained without supervised labels generated by exhaustive search. For each mini-batch, the loss function is defined as the negative average sum spectral efficiency, i.e.,
\begin{equation}
\mathcal{L}_{\rm loss}
=
-\frac{1}{B_{\rm mb}}\sum_{b=1}^{B_{\rm mb}}R_b,
\end{equation}
where $B_{\rm mb}$ is the mini-batch size and $R_b$ is calculated by \eqref{eq:se} for the $b$-th channel realization. The Gumbel-Softmax temperature is annealed independently of the learning-rate schedule; Adam starts from $10^{-4}$ and the learning rate is multiplied by $0.5$ every $10$ epochs, without a warmup stage. This objective directly matches the communication performance metric and enables joint optimization of EM precoding, analog precoding, and digital precoding.

\section{Simulation Results}

Unless otherwise stated, we consider $N_{\rm t}=32$ transmit antennas arranged as a planar array with $N_x=4$ antennas along the horizontal dimension and $N_z=8$ antennas along the vertical dimension, $N_{\rm RF}=8$ RF chains, the number of users $U=2$, $N_{\rm c}=60$ adopted OFDM subcarriers, and $L_u=6$ paths for all users. The transmit power is varied from $10$ to $50$ dBm and fixed at $50$ dBm where otherwise indicated. The $4\times8$ array is partitioned into eight disjoint contiguous $2\times2$ subarrays, each comprising two antennas along both the horizontal and vertical dimensions and connected to one RF chain; the subarrays are formed by grouping adjacent horizontal and vertical antenna indices. Unless otherwise noted, Tri-PNet uses this sub-connected RF architecture. In the fully-connected Tri-PNet comparison below, the same $N_{\rm RF}=8$ RF chains are connected to all $N_{\rm t}=32$ antennas. The OFDM numerology employs a $2048$-point fast Fourier transform and a subcarrier spacing of $\Delta f=15$ kHz in the unified wideband geometric channel model in \eqref{eq:unified_channel}. The carrier frequency is $f_{\rm c}=2.44$ GHz, and the phase quantizer in \eqref{eq:frf_quant} uses $b=4$ bits. For the non-regular mode  in Fig.~\ref{fig:rmm_patterns}, each antenna selects one of $N_{\rm p}^{\mathrm{NReg}}=4$ candidate patterns. For the 3GPP TR 38.901 mode with HPBW $=90^\circ$ in Fig.~\ref{fig:rpr_patterns}, each antenna selects one of $N_{\rm p}^{\mathrm{3GPP}}=9$ patterns generated from elevation and azimuth samples at $-60^\circ$, $0^\circ$, and $60^\circ$ on both axes. Three independent datasets, each containing $N_{\rm set}=10000$ channel samples, are generated; two are used for training and the remaining dataset is used for testing. The core training configuration is summarized in Table~\ref{tab:training_hyperparameters}.

\begin{table}[t]
\centering
\caption{Training hyperparameters.}
\label{tab:training_hyperparameters}
\footnotesize
\begin{tabular}{M{1.58in}|M{1.50in}}
\hline
\textbf{Hyperparameter} & \textbf{Value} \\
\hline
Optimizer & Adam \\
\hline
Training epochs & $50$ \\
\hline
Mini-batch size & $512$ \\
\hline
Initial learning rate & $10^{-4}$ \\
\hline
Learning-rate schedule & $\times0.5$ every $10$ epochs \\
\hline
Gumbel temperature & $\max\{0.1,2\exp(-0.03e)\}$ \\
\hline
\end{tabular}
\end{table}

\subsection{Spectral Efficiency Evaluation}
\responsechange{Throughout the following simulations, ``spectral efficiency'' denotes the subcarrier-averaged sum spectral efficiency }\responsechange{\currentchange{$R$ in \eqref{eq:se}. The sum over users}}\responsechange{ is averaged over subcarriers, without division by $U$.} For a fair comparison, all schemes use the same channel realizations, candidate pattern codebook, and transmit-power normalization. The EM precoding baselines use the same sub-connected RF architecture and phase-resolution constraint as Tri-PNet, whereas the fully-digital MIMO and fully-connected hybrid analog-digital MIMO benchmarks are configured as described below. The compared schemes are summarized as follows.
\begin{itemize}
\item \textbf{Tri-PNet (EM precoding with sub-connected hybrid analog-digital precoding)}\currentchange{.} The complete proposed framework jointly infers the antenna-wise EM precoder, the frequency-independent analog precoder, and the subcarrier-wise digital precoders. Its RPSNet and HPNet are trained together so that the EM precoder is optimized for the subsequent multi-user hybrid-precoding stage rather than only for antenna gain.
\item \textbf{Tri-PNet (EM precoding with fully-connected hybrid analog-digital precoding)}\currentchange{.} This variant retains the proposed EM precoder and Tri-PNet learning architecture, but replaces the sub-connected RF network with a fully-connected $N_{\rm t}\times N_{\rm RF}$ analog precoder. Thus, each of the $N_{\rm RF}=8$ RF chains is connected to all $N_{\rm t}=32$ antennas.
\item \textbf{PR-HBFNet \cite{prh} (EM precoding with sub-connected hybrid analog-digital precoding)}\currentchange{.} This baseline is the original cascaded learning framework composed of a Transformer-based radiation pattern reconfigurable network (PRN) and a model-driven hybrid beamforming network (HBN). The PRN selects one radiation pattern for the EM precoding of each antenna, while the HBN predicts residual analog and digital refinements over SVD-based initializations.
\item \textbf{RPSNet (EM precoding) + HBN \cite{prh} (Sub-connected hybrid analog-digital precoding)}\currentchange{.} The original PRN is replaced by the proposed Conformer-based RPSNet, whereas the HBN is retained. This configuration isolates the improvement obtained from local and global frequency-domain modeling in the proposed EM precoding.
\item \textbf{PRN \cite{prh}  + HPNet (Sub-connected hybrid analog-digital precoding)}\currentchange{.} The original Transformer-based PRN is combined with the proposed HPNet. It evaluates the gain provided by cross-attention analog precoding and dual-path digital precoding without changing the baseline EM-precoding network.
\item \textbf{Greedy selection \cite{rmm} (EM precoding) + HPNet or HBN \cite{prh} (Sub-connected hybrid analog-digital precoding)}\currentchange{.} The greedy approach updates one antenna at a time, evaluates every candidate pattern while holding the other antenna patterns fixed, and retains the pattern yielding the largest spectral efficiency. Coupling it separately with HPNet and HBN distinguishes the effect of iterative pattern search from that of the hybrid-precoding network.
\item \textbf{Random selection (EM precoding) + HPNet or HBN  \cite{prh} (Sub-connected hybrid analog-digital precoding)}\currentchange{.} Each antenna's radiation pattern is randomly sampled from the candidate codebook. The EM precoded channel is subsequently processed by HPNet or HBN, providing a low-complexity reference that does not exploit EM-domain adaptation.
\item \textbf{Hybrid precoding with fixed radiation pattern  \cite{rmm} (Sub-connected hybrid MIMO with fixed Type-0 radiation pattern)}\currentchange{.} All antennas are fixed to the default radiation pattern, corresponding to the legacy Type-0 initialization in \cite{rmm}. The analog precoder is obtained from singular-vector aggregation, followed by a ZF digital precoder; therefore, this baseline contains analog and digital precoding without EM precoding.
\item \textbf{WMMSE \cite{wmmse} (Fully-digital MIMO without EM precoding)}\currentchange{.} Eight antennas are fixed to the Type-0 radiation pattern, and a fully-digital weighted minimum mean-square-error (WMMSE) precoder is independently optimized on every subcarrier. It employs one RF chain per active transmit antenna and therefore uses eight RF chains for eight active antennas; it has neither an analog precoder nor an adaptive EM precoder.
\item \textbf{Principal-component analysis \cite{pca} (PCA-based fully-connected hybrid precoding)}\currentchange{.} All antennas are fixed to the Type-0 radiation pattern. Under a fully-connected hybrid MIMO architecture, a frequency-flat analog precoder is obtained by principal-component analysis of the subcarrier-wise fully-digital precoders, followed by subcarrier-wise digital precoding.
\item \textbf{Manifold optimization-based alternating minimization \cite{mo} (MO-AltMin-based fully-connected hybrid precoding)}\currentchange{.} All antennas are fixed to the default Type-0 radiation pattern. Under a fully-connected hybrid MIMO architecture, MO-AltMin refines the constant-modulus analog precoder on its phase manifold, followed by subcarrier-wise digital precoding.
\end{itemize}

We first evaluate the above schemes under the non-regular mode using the four Huawei-provided basis radiation patterns \cite{rmm}. According to~\eqref{eq:EM_matrix}, we use the gain function $G_p^{(\mathrm{EM-NReg})}(\phi_{u,\ell},\theta_{u,\ell})$ with four candidate radiation patterns in the unified channel model. 
We then evaluate Tri-PNet under the 3GPP TR 38.901 mode adapted from \cite{rpr}. In this setting, the normalized gain function $G_p^{(\mathrm{EM-3GPP})}(\phi_{u,\ell},\theta_{u,\ell})$ is constructed with nine candidate radiation patterns under the HPBW $=90^\circ$ setting and is embedded into each multipath component in the unified channel model. 

\begin{figure*}[t]
\centering
\includegraphics[width=0.94\textwidth]{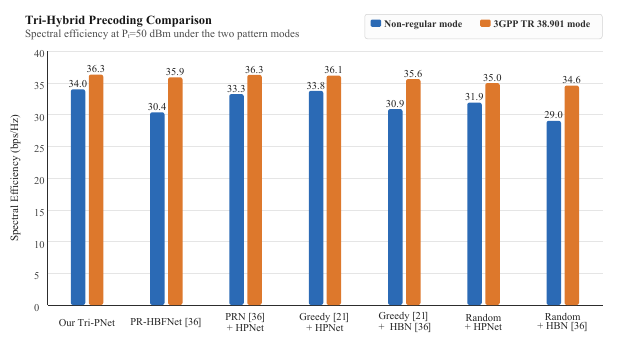}
\caption{Performance comparison of tri-hybrid precoding using different schemes under the non-regular and 3GPP TR 38.901 modes at $P_{\rm t}=50$ dBm.}
\label{fig:pattern_method_group}
\end{figure*}

Fig.~\ref{fig:pattern_method_group} gives a compact comparison of the tri-hybrid precoding under the two radiation-pattern modes. In addition to the complete PR-HBFNet \cite{prh}, the grouped bars combine its PRN with the proposed HPNet and combine greedy or random selection with either HPNet or the original HBN. The complementary RPSNet + HBN configuration is reported in the complexity tables. Replacing only one component improves performance to different degrees, whereas the complete Tri-PNet achieves the highest spectral efficiency. This trend indicates that the EM precoder and hybrid analog-digital precoder should be optimized together, because the selected pattern changes the channel observed by the analog and digital precoding stages.

\begin{figure*}[t]
\centering
\includegraphics[width=0.98\textwidth]{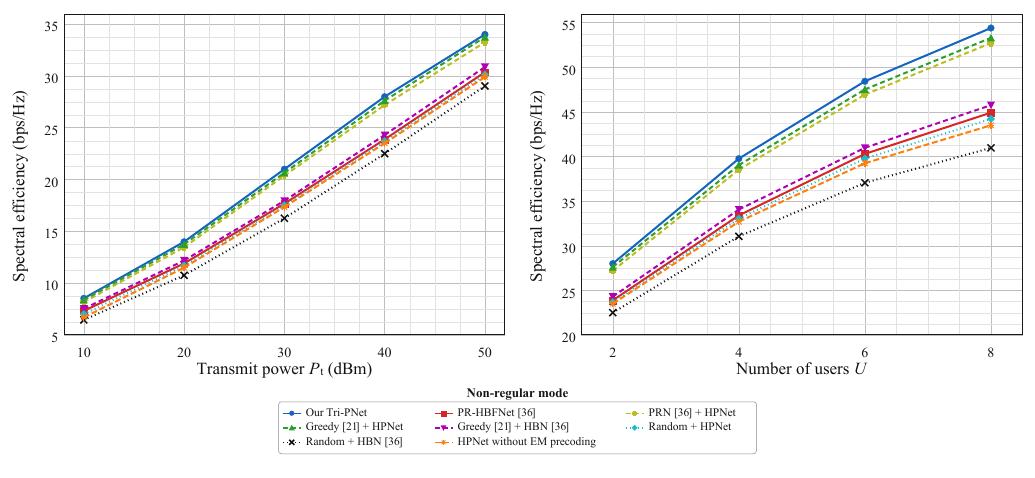}
\caption{Spectral-efficiency evaluation under the non-regular radiation-pattern mode. \responsechange{The left panel varies transmit power with $U=2$, and the right panel varies the number of users with $P_{\rm t}=50$ dBm and $U\in\{2,4,6,8\}$.}}
\label{fig:rmm_se_pt}
\label{fig:rmm_se_user}
\label{fig:rmm_se}
\vspace{-1mm}
\end{figure*}

Under the non-regular mode, the proposed Tri-PNet consistently achieves the highest spectral efficiency as the transmit power increases and as more users are served, as shown in Fig.~\ref{fig:rmm_se}. The gain over various baselines demonstrates the benefit of jointly learning EM precoding, analog precoding, and digital precoding. In particular, the gap between Tri-PNet and the random-selection curves confirms that the learned EM precoding exploits instantaneous CSI.

\begin{figure*}[t]
\centering
\includegraphics[width=0.98\textwidth]{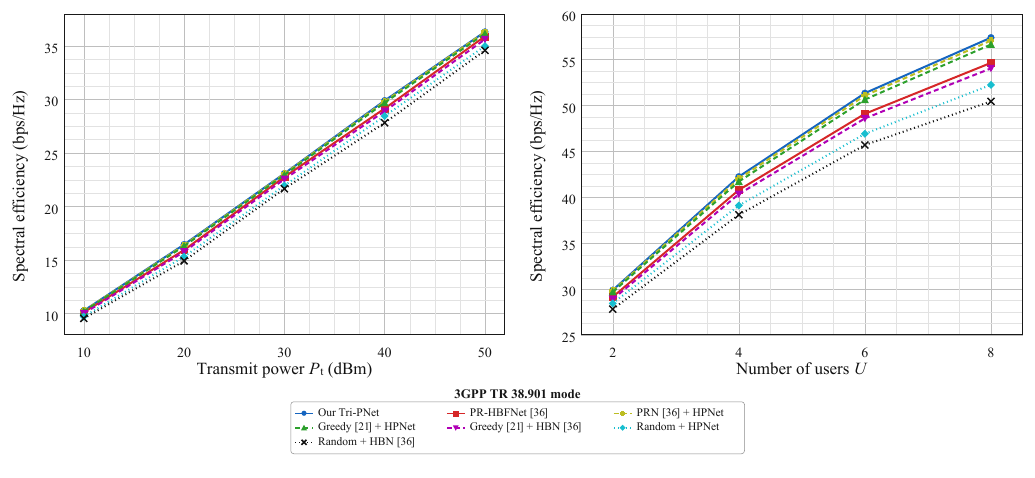}
\caption{\responsechange{Spectral-efficiency evaluation under the 3GPP TR 38.901 radiation-pattern mode. The left panel varies transmit power with $U=2$, and the right panel varies the number of users with $P_{\rm t}=50$ dBm and $U\in\{2,4,6,8\}$.}}
\label{fig:rpr_se_pt}
\label{fig:rpr_se_user}
\label{fig:rpr_se}
\vspace{-1mm}
\end{figure*}

The 3GPP TR 38.901 results in Fig.~\ref{fig:rpr_se} exhibit similar overall performance trends. Tri-PNet remains competitive with the greedy radiation-pattern selection benchmark for EM precoding and outperforms the learning-based and fixed-pattern alternatives over the tested power and user ranges. \responsechange{These results evaluate the framework under the two specified pattern families, without establishing transfer of a fixed trained model between codebooks.} As the number of users grows, the persistent advantage of Tri-PNet further supports its ability to use cross-user and cross-subcarrier structure when multi-user interference becomes more severe.

\begin{figure*}[t]
\centering
\includegraphics[width=0.98\textwidth]{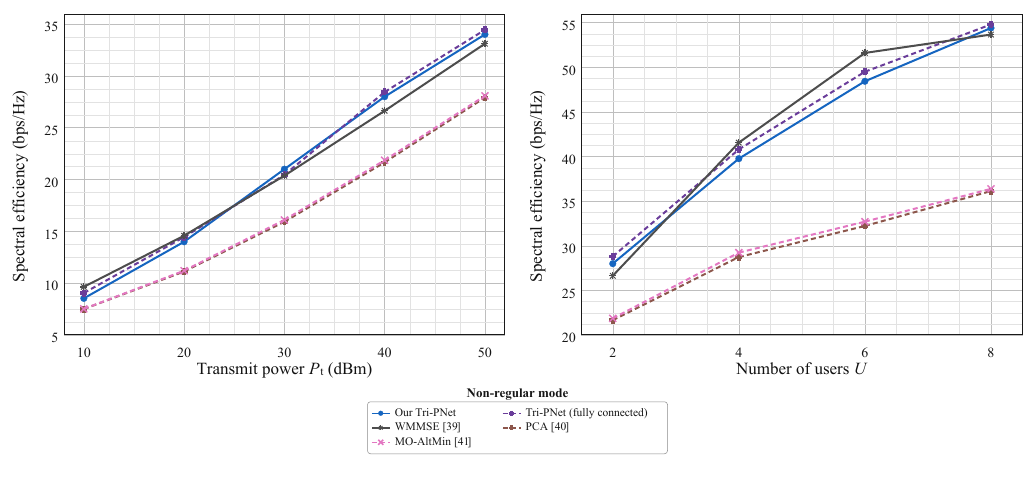}
\caption{Comparison of the proposed sub-connected Tri-PNet, fully-connected Tri-PNet, fully-digital WMMSE, and fixed-Type-0 fully-connected PCA and manifold optimization-based alternating minimization (MO-AltMin) precoding under the non-regular radiation-pattern mode. The fully-connected Tri-PNet uses $N_{\rm RF}=8$ RF chains connected to all $N_{\rm t}=32$ antennas, whereas fully-digital WMMSE uses eight RF chains for eight active antennas. The left and right panels show the spectral efficiency versus transmit power and number of users, respectively. Here the user numbers are $U\in\{2,4,6,8\}$.}
\label{fig:se_full_digital}
\vspace{-1mm}
\end{figure*}

Fig.~\ref{fig:se_full_digital} primarily compares the proposed sub-connected Tri-PNet with conventional fully-connected or sub-connected hybrid precoding baselines using a fixed Type-0 radiation pattern, including PCA and MO-AltMin, as well as the fully-digital WMMSE benchmark. Unlike the conventional hybrid schemes, Tri-PNet jointly exploits EM precoding, sub-connected analog precoding, and digital precoding, thereby adapting the effective channel before hybrid beamforming. The performance gains over PCA and MO-AltMin demonstrate that EM-domain adaptation provides an additional degree of freedom beyond conventional analog--digital hybrid precoding. \responsechange{For the common eight-RF-chain budget, WMMSE uses eight active antennas with fixed Type-0 patterns, whereas Tri-PNet uses 32 antennas and adaptive pattern selection. This comparison is not at equal aperture, and the plotted WMMSE reference is not a universal fully-digital upper bound.} Meanwhile, to quantify the performance cost imposed by the sub-connected RF constraint, we also report the fully-connected Tri-PNet as an architecture-aware reference. The gap between the two Tri-PNet variants characterizes the spectral-efficiency loss incurred by restricting each RF chain to a local subarray, whereas the advantage of sub-connected Tri-PNet over the conventional PCA-based and MO-AltMin-based fully-connected hybrid precoding with fixed Type-0 radiation pattern baselines verifies that the proposed tri-hybrid design remains effective under the sub-connected architecture.

\subsection{Spectral Efficiency Versus HPBW}
\begin{figure}[t]
\centering
\includegraphics[width=0.98\linewidth]{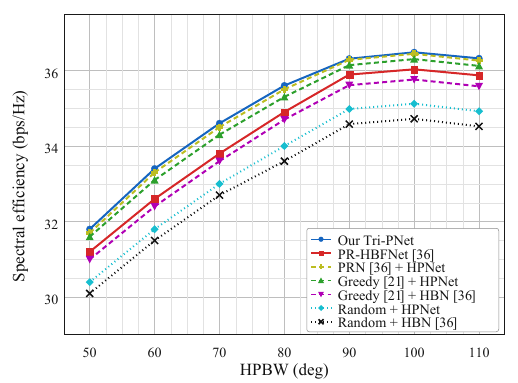}
\caption{Spectral efficiency versus HPBW under the 3GPP TR 38.901 pattern mode at $P_{\rm t}=50$ dBm.}
\label{fig:se_hpbw}
\end{figure}

\responsechange{To examine the sensitivity to the radiation-beamwidth setting, we vary the HPBW of the 3GPP TR 38.901 pattern family while keeping the transmit power at $P_{\rm t}=50$ dBm. The varied HPBW describes each candidate radiation pattern of an individual antenna element, rather than the overall precoded array beamwidth.} The evaluated HPBW values range from $50^\circ$ to $110^\circ$. For each value, the 3GPP TR 38.901 mode is constructed from the nine direction-dependent patterns generated by the three elevation-axis and three azimuth-axis samples, and all schemes use the same channel realizations and hybrid-precoding settings. As shown in Fig.~\ref{fig:se_hpbw}, spectral efficiency first increases with HPBW because wider patterns improve angular coverage, reaches its maximum near $100^\circ$, and then decreases slightly at $110^\circ$ because excessive beam overlap reduces directional discrimination and weakens multi-user interference suppression. Tri-PNet remains the best-performing method over the complete range, while the $90^\circ$ point coincides with the 3GPP TR 38.901 setting used in the previous simulations. The consistent ordering indicates that the learned tri-hybrid precoding remains beneficial across different HPBW settings.

\subsection{Robustness Under Imperfect CSI}
\currentchange{To evaluate robustness under imperfect CSI, possibly caused by channel estimation error \cite{isac_cs} or CSI aging, we follow} the normalized channel perturbation error (NCPE) metric used in \cite{tbh}. \responsechange{Specifically, $[\mathbf{h}^{\rm EM}[q]]_{p,u}=[h_{p,q,u,1},\ldots,h_{p,q,u,N_{\rm t}}]^{\rm T}\in\mathbb{C}^{N_{\rm t}}$ collects the entries $[\mathcal{H}^{\rm EM}]_{p,q,u,:}$ of the candidate CSI tensor in (14). It is defined before pattern selection and differs from the selected vector $\mathbf{h}_{u}^{\rm EM}[q]$ in \eqref{eq:selected_channel}.} For both radiation-pattern modes, the perturbed candidate channel is modeled as $[\mathbf{h}_{\rm per}^{\rm EM}[q]]_{p,u}=[\mathbf{h}^{\rm EM}[q]]_{p,u}+[\mathbf{n}_{\rm per}^{\rm EM}[q]]_{p,u}$, where $[\mathbf{n}_{\rm per}^{\rm EM}[q]]_{p,u}\sim\mathcal{CN}(\mathbf{0},\sigma_{\rm per}^{2}\mathbf{I}_{N_{\rm t}})$ is complex Gaussian perturbation noise, and the NCPE is defined as \({\rm NCPE}^{\rm EM}=\frac{\sum_{p=1}^{N_{\rm p}}\sum_{u=1}^{U}\sum_{q=1}^{N_{\rm c}}\left\|[\mathbf{h}^{\rm EM}[q]]_{p,u}-[\mathbf{h}_{\rm per}^{\rm EM}[q]]_{p,u}\right\|_2^2}{\sum_{p=1}^{N_{\rm p}}\sum_{u=1}^{U}\sum_{q=1}^{N_{\rm c}}\left\|[\mathbf{h}^{\rm EM}[q]]_{p,u}\right\|_2^2}\).

\begin{figure}[t]
\centering
\includegraphics[width=0.92\linewidth]{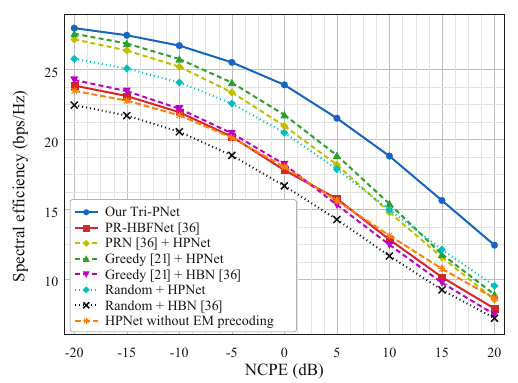}
\caption{Spectral efficiency versus NCPE under the non-regular mode. The NCPE values range from $-20$ to $20$ dB.}
\label{fig:rmm_se_ncpe}
\end{figure}

Fig.~\ref{fig:rmm_se_ncpe} shows the spectral efficiency under channel perturbation for the non-regular mode. As NCPE increases, the channel features observed by the tri-hybrid precoding become less accurate, and all schemes experience performance degradation. Tri-PNet remains above baselines because the Conformer-based RPSNet extracts local and global frequency-domain features, while the HPNet uses model-driven analog and digital precoding priors to reduce sensitivity to noisy CSI. The consistent gain confirms that the proposed network preserves the benefit of coupled tri-hybrid precoding under imperfect CSI.

\begin{figure}[t]
\centering
\includegraphics[width=0.92\linewidth]{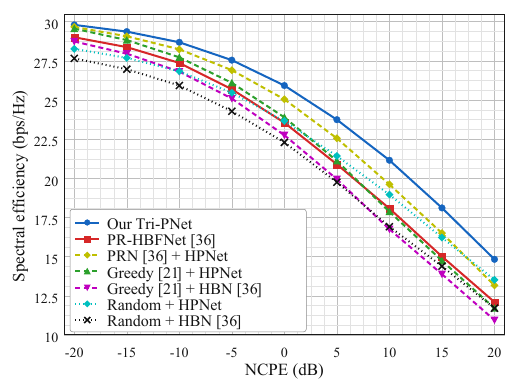}
\caption{Spectral efficiency versus NCPE under the 3GPP TR 38.901 mode. The NCPE values range from $-20$ to $20$ dB.}
\label{fig:rpr_se_ncpe}
\end{figure}

Fig.~\ref{fig:rpr_se_ncpe} evaluates the same robustness metric for the 3GPP TR 38.901 mode. Since the candidate patterns are direction-dependent, perturbations in CSI directly affect tri-hybrid precoding design. Tri-PNet provides a clear gain over PR-HBFNet \cite{prh} and greedy selection \cite{rmm} throughout the tested perturbation range. This result shows that the learned tri-hybrid precoding preserves robustness without performing an online pattern search and continues to exploit the direction-dependent pattern diversity when the CSI is imperfect.

\begin{table}[t]
\centering
\caption{Parameter number and FLOPs under the non-regular mode.}
\label{tab:complexity_rmm}
\begin{tabular}{|M{1.45in}|M{0.72in}|M{0.72in}|}
\hline
Configuration & Parameters & FLOPs\\
\hline
PR-HBFNet \cite{prh} & 8.65M & 481.03M\\
\hline
RPSNet + HBN \cite{prh} & 10.83M & 619.79M\\
\hline
PRN \cite{prh} + HPNet & 11.56M & 696.80M\\
\hline
Our Tri-PNet & 13.74M & 835.55M\\
\hline
Greedy \cite{rmm} + HPNet & 7.98M & 544.75M\\
\hline
\end{tabular}
\end{table}

\begin{table}[t]
\centering
\caption{Parameter number and FLOPs under the 3GPP TR 38.901 mode.}
\label{tab:complexity_rpr}
\begin{tabular}{|M{1.45in}|M{0.72in}|M{0.72in}|}
\hline
Configuration & Parameters & FLOPs\\
\hline
PR-HBFNet \cite{prh} & 13.12M & 487.92M\\
\hline
RPSNet + HBN \cite{prh} & 15.30M & 626.67M\\
\hline
PRN \cite{prh} + HPNet & 16.03M & 703.68M\\
\hline
Our Tri-PNet & 18.21M & 842.43M\\
\hline
Greedy \cite{rmm} + HPNet & 7.98M & 555.24M\\
\hline
\end{tabular}
\end{table}
\subsection{Complexity Analysis}
The computational complexity under the two radiation-pattern modes is summarized in Tables~\ref{tab:complexity_rmm} and~\ref{tab:complexity_rpr} in terms of the number of trainable parameters and FLOPs per inference. The 3GPP TR 38.901 mode contains more candidate radiation patterns than the non-regular mode, resulting in higher RPSNet-related parameter and FLOP counts. In contrast, the HPNet-related complexity remains unchanged because HPNet operates on the selected channels $\{\mathbf{H}^{\rm EM}[q]\}^{N_{\rm c}}_{q=1}$ and is independent of the number and physical source of the candidate radiation patterns.

For greedy pattern selection, $N_{\rm t}$ antennas are updated sequentially and $N_{\rm p}$ candidate patterns are evaluated at each update. Accounting for the associated hybrid-precoder calculation over $N_{\rm c}$ subcarriers, its online complexity scales approximately as $\mathcal{O}(N_{\rm t}N_{\rm p}N_{\rm c}U N_{\rm RF}^{2})$. In contrast, after offline training, Tri-PNet jointly produces its radiation pattern selection for EM precoding and hybrid precoders through a single forward inference. Although Tri-PNet introduces more trainable parameters than PR-HBFNet~\cite{prh}, it avoids iterative pattern search~\cite{rmm} while providing the spectral-efficiency gains demonstrated in the preceding simulations.

\section{Conclusion}
This paper develops Tri-PNet, a Conformer-based tri-hybrid precoding framework for wideband multi-user downlink systems. A unified radiation-pattern-aware wideband channel model is considered, where the non-regular pixel-antenna mode and the 3GPP TR 38.901 mode are represented through different radiation-pattern gain matrices. Based on this channel model, Tri-PNet jointly learns EM precoding, analog precoding, and digital precoding from sample-wise channel tensors. By combining the Conformer-based RPSNet with cross-attention analog precoding and dual-path digital precoding in the HPNet, Tri-PNet captures the coupling among users, antennas, subcarriers, and radiation pattern modes. 
Simulation results demonstrate that Tri-PNet improves spectral efficiency over various baseline schemes while approaching the greedy benchmark with substantially lower online complexity by replacing iterative pattern search with a single forward inference. The results further reveal a non-monotonic relationship between the HPBW and spectral efficiency, with the best performance observed near $100^\circ$ under the 3GPP TR 38.901 radiation-pattern mode, highlighting the importance of properly selecting the radiation-pattern beamwidth. Future work will extend Tri-PNet to multi-cell scenarios and investigate its performance under more realistic spatially consistent channel models.

\balance

\end{document}